\documentclass[%twocolumn,
aps,nofootinbib,
showpacs,showkeys,preprint
tightenlines,preprintnumbers,] {revtex4}

\usepackage{epsf,epsfig,subfigure,graphicx,amsmath,amssymb %,mathtools
}
\usepackage{color}
\newcommand{\dis}[1]{\begin{equation}\begin{split}#1\end{split}\end{equation}}
\newcommand{\ie}{{\it i.e.~}}

 \newcommand{\dell}{\delta_{\rm PMNS}}
\newcommand{\delq}{\delta_{\rm CKM}}

\newcommand{\Qem}{Q_{\rm em}}

\def\sw0{{$\sin^2\theta_W^0$}}

\newcommand{\Z}{{\bf Z}}

\def\E6{{\rm E_6}}

\def\EE8{{\rm E_8\times E_8'}}

\def\one{{\bf 1}}

\def\five{{\bf 5}}
\def\ten{{\bf 10}}

\def\fiveb{\overline{\bf 5}}

\begin{document}

\draft

\title{\Large\bf Unifying CP violations of quark and lepton sectors}

\author{ Jihn E. Kim$^{(a,b,c)}$ and Soonkeon Nam$^{(a)}$ }
 
\address{ 
 $^{(a)\,}$Department of Physics, Kyung Hee University, 26 Gyungheedaero, Dongdaemun-Gu, Seoul 02447, Republic of Korea,  \\
$^{(b)\,}$Center for Axion and Precision Physics Research (IBS),
  291 Daehakro, Yuseong-Gu, Daejeon 34141, Republic of Korea,\\ 
$^{(c)\,}$Department of Physics, Seoul National University, 1 Gwanakro, Gwanak-Gu, Seoul 08826, Republic of Korea 
}

\begin{abstract} 
A preliminary determination of the Dirac phase in the PMNS matrix is $\dell\approx -\frac{\pi}{2}$. A rather accurately determined Jarlskog invariant $J$ in the CKM matrix is close to the maximum. Since the phases in the CKM and PMNS matrices will be accurately determined in the future,  it is an interesting problem to relate these two phases. This can be achieved in a families-unified grand unification if the weak CP violation is introduced spontaneously {\it \`a la} Froggatt and Nielsen at a high energy scale, where only one meaningful Dirac CP phase appears.   
 
\end{abstract}
\maketitle

%%%%%%%%%%%%%%%%%%%%%%%%%%%%%%%%%%%%%%%%%%%%%%%%%%%%
%%%%%%%%%%%%%%%%%%%%%%%%%%%%%%%%%%%%%%%%%%%%%%%%%%%%

\section{Introduction}\label{sec:Introduction}
   
At present, the real angles of the Cabibbo-Kobayashi-Maskawa(CKM) matrix is rather accurately determined \cite{PDG15}, which makes it possible to pin down the invariant phase $\delq$ into three possibilities $\alpha,\beta$, and $\gamma$ of the unitarity triangle \cite{CKMUGUTF15}. The physically observable CP magnitude is the Jarlskog determinant $J$ \cite{Jarlskog85} which can be expressed as $J=(\rm real~angles)\cdot\sin\delq$. Depending on the parametrization, $\delq$ can be $\alpha,\beta,$ or $\gamma$. The maximality of $J$ is a different concept from the maximality of the phase $\delq$. The maximality of phase is $\delq\simeq \pm\frac{\pi}{2}$. Even though $\delq\ne \pm\frac{\pi}{2}$, $J$ can be maximal in the vicinity of a given $\delq$, which can be checked by varying the real angles together with $\delq$ within the experimentally allowed bounds \cite{CKMUGUTF15}. $\delq$  is close to the maximum 90 degrees   in the parametrization suggested by Kim and  Seo (KS) \cite{SeoPRD11,KimJKPS15} and Kobayashi and Maskawa(KM) \cite{KM73,CKMUGUTF15}. The Particle Data Group(PDG) compilation of the invariant phase is $\alpha=(85.4^{+3.9}_{-3.8})^{\rm o}$ \cite{CKMPDG15}. Thus,  $\delq=\alpha$ in the KS and KM parametrizations shows that $J$ is close to maximum. The same maximality of $J$ is also drawn from the Chau-Keung(CK) parametrization where $\delq=\gamma$ \cite{CKMUGUTF15}.  

In this paper, we use the  KS parametrization as an explicit example  \cite{SeoPRD11} where the only complex number  in the CKM matrix \cite{Cabibbo63,KM73} is the invariant Jarlskog phase itself \cite{Jarlskog85}. Here, the phase is multiplied to the (small O($|\lambda^3|$)) whole element, $V_{\rm CKM\,(31)}$, which makes it possible to appreciate the weak CP violation from $V_{\rm CKM}$ itself. For the Pontecorvo-Maki-Nakagawa-Sakada(PMNS) matrix \cite{PMNS}, we have already entered into an era of determining a Dirac phase $\dell$ modified by some Majorana phases, with a preliminary result close to $\mp 90$  degrees at a  $2\,\sigma$ level \cite{OyamaPlanck15}. Therefore, it is timely to ask a question now whether one can relate $\dell$ and $\delq$ or not. 
To relate the CKM and  PMNS phases, one can consider using a  grand unified theory(GUT) which unifies quarks and leptons with a suitable scheme unifying families \cite{Georgi79,KimJHEP15}.   

Most family unification models assume a factor group $G_f$ in addition to the Standard Model(SM) or GUT,  where continuous symmetries  such as SU(2) \cite{Zee79}, SU(3) \cite{KingRoss01}, or U(1)'s \cite{FN79,Chun01}, and discrete symmetries such as $S_3$ \cite{S3}, $A_4$ \cite{Ma01}, $\Delta_{96}$ \cite{King14}, $\Z_{12}$ \cite{Dodeca} for  $G_f$ have been considered. A full unification of GUT families in the sense that the couplings of the family group are unified with the three gauge couplings of the SM is by unifying the families in a simple gauge group based on SU($N$) \cite{Georgi79}. Along this line, one of the authors has  recently suggested a families-unification based on anti-SU(7)  GUT, SU(7)$\times$U(1) \cite{KimJHEP15}, which in fact unifies family couplings with three gauge couplings of the SM. String derived anti-SU(7) \cite{KimJHEP15} has a merit in that it is free from gauge anomalies and from the gravity spoil of some discrete symmetries  \cite{KraussWilczek, Ibanez92, Barr92, KimPRL13}. In this paper, however, we discuss at the field theory level of the SM. 

We need a true unification of GUT families. Even that requirement is used only when we argue for the possibility of $\delq\simeq\pm\dell $  based on the assumption  that the Dirac phases in the CKM and PMNS matrices originates from the {\it spontaneous CP violation} mechanism \cite{Lee73} at a high energy scale {\it \`a la} Froggatt and Nielsen \cite{FN79}.  

If one allows completely general complex Yukawa couplings in the quark and lepton sectors, one cannot relate $\dell$ and $\delq$. 
If all Yukawa couplings are real, the weak CP violation must be introduced by unremovable complex vacuum expectation values(VEVs). To have a relation without any other parameters, such as in the relation $\dell=\pm \delq$, only one phase must be introduced in the whole theory such as in the unification of GUT families. To mimick the KM weak CP at low energy, the complex VEV must be that of a SM singlet \cite{Nelson84} as performed in \cite{KimMax11}. 

In the short section Sec. \ref{sec:CKMPMNS}, we define  the CKM and PMNS matrices. Section \ref{subsec:Yukawa} is the main part of the paper, where the diagonalization of mass matrices and parametrization of the CKM and PMNS matrices are discussed. In Sec. \ref{sec:CCKMphase}, we present the diagonalzation mechanisms of $M^{(u)}$ and  $M^{(\nu)}$, needed for relating $\delq$ and $\dell$, and present a relation in the anti-SU(7) model.  Sec. \ref{sec:Conclusion} is a conclusion.

%%%%%%%%%%%%%%%%%%%
\section{The CKM and PMNS matrices}
\label{sec:CKMPMNS}
%%%%%%%%%%%%%%%%%%%
 
 In this short section, we define the CKM and PMNS matrices.
Let the quark and lepton representations of the SM be
\dis{
q_{a\,L}=\begin{pmatrix} u_a\\ d_a\end{pmatrix}_L,~u^c_{a\,L},~ d^c_{a\,L}; ~~ \ell_{a\,L}=\begin{pmatrix} \nu_a\\ e_a\end{pmatrix}_L, 
~ e^+_{a\,L},~ N_{a\,L},~a=1,2,3,\label{eq:OneSMfamily}
}
where $a$ is the family indices in the weak eigenstates.
The $(ij)$ element of the CKM matrix is defined as the $W_\mu^+$ boson coupling to the current $\bar{u}^{({\rm mass}\,i)}_L\gamma^\mu {d}^{({\rm mass}\,j)}_L\,W^{+}_\mu$ where ${u}^{({\rm mass}\,i)}$ and ${d}^{({\rm mass}\,i)}$ are the mass eigenstates,
\dis{
&{u}^{({\rm mass}\,1)}=u,~{u}^{({\rm mass}\,2)}=c,~{u}^{({\rm mass}\,3)}=t,\\ &{d}^{({\rm mass}\,1)}=d,\,~{d}^{({\rm mass}\,2)}=s,~{d}^{({\rm mass}\,3)}=b.\label{eq:SMfields}
}
Choosing the mass eigenstate $d$ quarks is quite general since this step is considered after diagonalizing the $d$-type quark masses, and below we will not touch upon the redefinition possibility of $d$-type quarks. This choice is useful in  the flipped-SU(5) GUT \cite{Barr82,DKN84} from string origin \cite{Ellis89,KimKyae07},  where $\Qem=-\frac13$ quarks and heavy neutrino $N$'s are grouped in $\ten$ on which we do not intend to question how they couple. Namely, we intend to discuss as much as possible without discussing the heavy neutrino sector.
Then,  the $(ij)$ element is the matrix element diagonalizing the weak states ${u}^{({\rm mass}\,i)}_L=\sum_{a} V_{ia}u^a_L$

%%%%%%%%%%%%%%%%%%
\begin{figure}[!t]
\begin{center}
\includegraphics[width=0.4\linewidth]{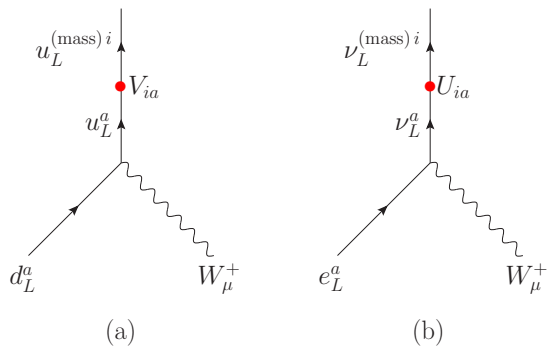}
\end{center}
\caption{The charged currents defining the CKM and PMNS matrices: (a) quarks, and (b) leptons.  In (a), to show the unitary transformation explicitly we represent the coupling $gV_{ia}$ as the red bullet and $u_L^a$ line, and similarly in (b). } \label{fig:CC}
\end{figure}
%%%%%%%%%%%%%%%%%%%%%

\dis{
V_{ij}^{\rm CKM}=V_{ia}.
}
This is depicted in Fig. \ref{fig:CC}\,(a). 
In the standard presentation of the KM model,
for the $W_\mu^+$ coupling we consider only the unitary matrix diagonalizing $q_L^a$ fields, together with the up-type and down-type quark phases. In our case, we already diagonalized down-type quarks and we consider the phases of right-handed up-type quarks  instead of the phases of left-handed down-type quarks. We draw the intermediate line $u_L^a$ in Fig. \ref{fig:CC}\,(a) to imply that it is related to  $u_R^b$ in the mass diagonalization process.
  Similarly, we define the  $W_\mu^+$ coupling to the lepton currents,  $\bar{\nu}^{({\rm mass}\,i)}_L\gamma^\mu {e}^{({\rm mass}\,j)}_L\,W^{+}_\mu$, where ${\nu}^{({\rm mass}\,i)}$ and ${e}^{({\rm mass}\,i)}$ are the mass eigenstates. This leptonic currents define the PMNS matrix. There is another reason to use the bases where charged leptons are mass eigenstates. It is because masses of $e,\mu,$ and $\tau$ are known accurately. Then, the PMNS marix is
\dis{
U_{ij}^{\rm PMNS}=U_{ia}.
}
Namely, the matrix diagonalizing the SM neutrinos is $U_{ia}$,
\dis{
{\nu}^{({\rm mass}\,i)}_L=\sum_{a} U_{ia}\nu^a_L,
}
and the PMNS matrix is depicted in Fig. \ref{fig:CC}\,(b).
 
%%%%%%%%%%%%%%%%%%%%%%%%%%%%%%%%%%%%%
\section{Yukawa couplings, masses, and spontaneous CP violation} \label{subsec:Yukawa}
  
Not to allow some complication on the flavor changing neutral current (FCNC) issue, let us introduce only one pair of BEH doublets: $H_d$ coupling to $d$-type quarks and $H_u$ coupling to $u$-type quarks, probably by a Peccei-Quinn symmetry \cite{Baer15}.  
 Masses for the charged leptons and  $\Qem=-\frac13$ quarks are arising from the Dirac Yukawa couplings $f_{ab}^{(e)}\, {\ell}_a e^+_bH_d$ and $f_{ab}^{(d)}\,q_a d^c_b H_d$, respectively. Let us   take $f_{ab}^{(e,d)}$ and  the VEV $\langle H_d^0\rangle$ as real values, and {\it diagonalize  the charged lepton  and $\Qem=-\frac13$ quark mass matrices}, without affecting the CP phase we would like to introduce. 
 One may encounter a situation where the Higgs field coupling to $d$-type  quarks develop complex VEVs, in which case equations take a bit more complicated forms. Not to clutter to this situation, we do not choose this vacuum.
Then, all CP violation effects are assumed to arise from the Yukawa couplings of $H_u$,  
\dis{
&f_{ab}^{(u)} q_a  u^c_b\, H_u,\\[0.2em] &f_{ab}^{(\nu)} {\ell}_a  N _b\, H_u.\label{eq:DiracUandNu}
 } 
Let us assume that $f_{ab}^{(u)} $ and $f_{ab}^{(\nu)} $ are real. Then, one $H_u$-type doublet cannot introduce  a weak CP violation spontaneously \cite{Lee73} even if it develops a complex VEV, say $\langle H_u^0\rangle=v_u \, e^{i \delta} $, since the phase appears as an overall one in the up-type quark mass matrix.  
So, we introduce phases in  $f_{ab}^{(u)} $ and $f_{ab}^{(\nu)} $ at a high energy scale by a complex VEV of {\it one} SM singlet field $X$ {\it \`a la} Froggatt and Nielsen \cite{FN79}.
Here, we allow $X$ couplings, not only one power but many different powers of $X$. Effectively, it amounts to intoducing many $H_u$'s but the FCNCs are suppressed by superheavy masses of $X$.
  By some symmetry structure of the theory, the $X$ coupling can be made flavor-dependent \cite{KimmaxCP}. 
Now, let us proceed in this scheme to relate $\delq$ and $\dell$.

\begin{figure}[!t]
\begin{center}
\includegraphics[width=0.55\linewidth]{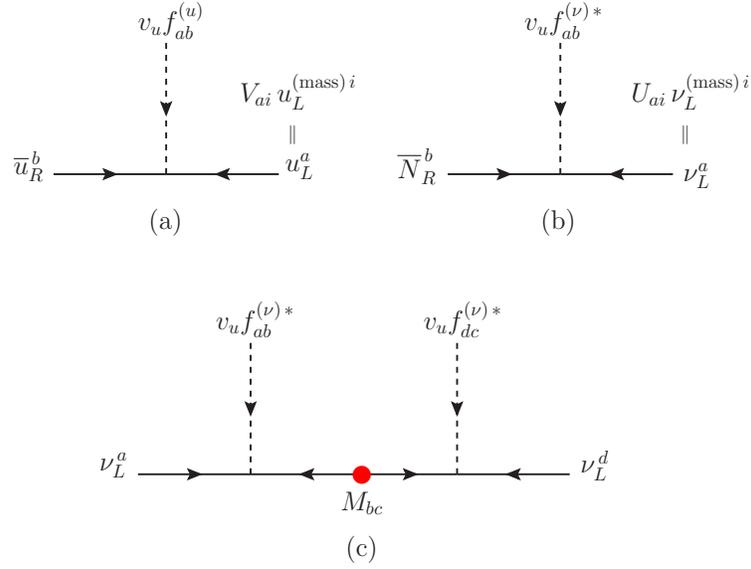}
\end{center}
\caption{The diagrams for masses of up-type quarks  and neutrinos: (a) the Dirac mass of $u^a$ , (b) the Dirac mass of $N$ and $\nu$, and (c) the seesaw mass of the SM neutrinos. The bullet in (c) is the Majorana mass $M_{bc}$ of heavy neutrinos $N^b$ and $N^c$.} \label{fig:MajNuMass}
\end{figure}

The Dirac Yukawa coupling (\ref{eq:DiracUandNu}) gives masses to both $\Qem=\frac23$ quarks and neutrinos. Let us take that $\langle H_u\rangle$ is real. If it were complex, its phase can be removed by redefining $\Qem= \frac23$ quark fields. Then, starting from the weak eigenstate bases, we obtain
\dis{
L^{(u)} &= \bar{u}_R^b f_{ab}^{(u)}\langle H_u^0\rangle {u}_L^a
+ \bar{u}_R^b f_{ba}^{(u)*}\langle H_u^0\rangle {u}_L^a\\
&=v_u\bar{u}_R^b f_{ab}^{(u)} (V^\dagger)_{ai}{u}^{({\rm mass}\,i)}_L
+v_u \bar{u}_R^b f_{ba}^{(u)*} (V^\dagger)_{ai}{u}^{({\rm mass}\,i)}_L 
\label{eq:uquarkMass}
}
and
\dis{
L^{(\nu)} &= \overline{N}_R^b f_{ab}^{(\nu)*}\langle \tilde{H}_u^0\rangle {\nu}_L^a =v_u\overline{N}_R^b f_{ab}^{(\nu)*}  (U^\dagger)_{ai}{\nu}^{({\rm mass}\,i)}_L \label{eq:nuMass}
}
where $ \tilde{H}_u\ =i\sigma_2 H_u^*=(H_u^{0\,*}, -H^-)^T$ with $\langle  {H}_u^{0 }\rangle=\langle \tilde{H}_u^{0\,*}\rangle=v_u$. In Eq. (\ref{eq:nuMass}),  the fact of only one chirality, say the left-handedness of the SM neutrinos is used. As commented above, the Yukawa couplings $f_{ab}^{(u)}$ and $f_{ab}^{(\nu)}$ can be complex  {\it \`a la} Froggatt and Nielsen. In Fig. \ref{fig:MajNuMass}, we visualize the mass terms of the up-type quarks and neutrinos.
For the neutrinos, the Type-I seesaw mechanism is used. 

%%%%%%%%%%%%%%%%%%
%%%%%%%%%%%%%%%%%%%%%

To relate the phases in the CKM and PMNS matrices, the phases of $f_{ab}^{(u)}$ and $f_{ab}^{(\nu)}$ must be related. Here, we need some model for family unification. 
As commented in Introduction, we use the top-down approach, \ie the model of the true unification of GUT families based on anti-SU(7) from string compactification.
In the family unification models from the bottom-up approach, one has to check the  vanishing of some anomalous terms via the discrete gauge symmetry \cite{KraussWilczek}, which is not a simple task. 

The sixteen chiral fields of the SM are grouped into $\ten, \fiveb$ and $\one$ of the  flipped-SU(5) spectrum (or anti-SU(5) \cite{DKN84}), contained in anti-SU(7) \cite{KimJHEP15}, as,
 \dis{
 \ten_a= \begin{pmatrix} |~  u_a~ | \\
  d^c_a ~~| ~~~~~  | ~N_a\\   |~  d_a~ |\end{pmatrix}_L,~
    \fiveb_a=   \begin{pmatrix} u^c_a\\ -- \\ \ell_a\end{pmatrix}_L ,~\one_a=e^+_{a\,L}, \label{eq:OneSU5family}
}
where bars separate different color  representations. Here, $\ell_a$ is the $a$-th lepton doublet,
\dis{
 \ell_a=\begin{pmatrix}\nu_a\\ e_a \end{pmatrix} .\nonumber
}
  Then, the couplings in Eq. (\ref{eq:uquarkMass},\ref{eq:nuMass}) are the same,  $F_{ab}  =f_{ab}^{(u)}=f_{ab}^{(\nu)}$. The relevant phases are read for the same order of family indices $f_{ab}$'s in  Eqs. (\ref{eq:uquarkMass}) and (\ref{eq:nuMass}). Thus, the phases in the quark and lepton charged current is made to contain  $ \delta$. The argument is presented in Sec. \ref{sec:CCKMphase}. 

The physically relevant quantity measurable experimentally is the Jarlskog determinant $J$. The quark sector $J_{\rm CKM}$  contains the Jarlskog phase $\delq$ and the lepton sector $J_{\rm PMNS}$  contains the Jarlskog phase $\dell$. Here, we suggest how these are related to $\delta$. The Jarlskog determinant $J$ can be expressed simply as $J = {\rm Im}\,V_{31}^*V_{22}^*V_{13}^*$ \cite{KimJKPS15}.   It has the form $J= ({\rm product ~of~real~CKM~angles})\cdot\sin\delq = {\rm Im}\,[({\rm product ~of~real~ CKM~angles})\,e^{i\delq}] $. The {\it invariant} Jarlskog phase  $\delq$ is determined  up to three classes, $\alpha,\beta$ and $\gamma$ of PDG, depending on the parametrization schemes \cite{CKMUGUTF15}.   Let us use
the simple KS form for the CKM matrix  \cite{SeoPRD11}
 \dis{
 V^{\rm KS}_{\rm CKM}&=\left(
\begin{array}{ccc}
c_1 & s_1c_3 & s_1s_3  \\
-c_2s_1 & e^{-i\delq}s_2s_3+c_1c_2c_3 & -e^{-i\delq}s_2c_3+c_1c_2s_3  \\
-e^{i\delq}s_1s_2 & -c_2s_3+c_1s_2c_3 e^{i\delq} & c_2c_3+c_1s_2s_3e^{i\delq} \\
\end{array}\right) \label{eq:KSCKM}
}
where  the real CKM angles are $c_i=\cos\theta_i,s_i=\sin\theta_i$ for $i=1,2,3$. 
One merit of the form (\ref{eq:KSCKM}) is that $e^{i\delq}$ is the {\it overall phase in the small element}, \ie in the (31) element  in $V^{\rm KS}_{\rm CKM}$. This makes it easy to glimpse the magnitude of the Jarlskog determinent. In comparison, note that $e^{i\delq}$ does not appear as an overall phase in the CK parametrization. 

Similarly, the KS form for the PMNS matrix can be written from Eq. (\ref{eq:KSCKM}) by replacing $\theta_i\to \Theta_i, s_i\to S_i, c_i\to C_i,$ and $\delq\to \dell$. Of course, the real angles $\theta_i$  and $\Theta_i$ are not identical, because the mass matrices to be diagonalized involve unrelated ingredient. 
We note that 
\dis{
&J_{\rm CKM}=-{\rm Im}\,[V_{31}V_{22}V_{13}]= c_1c_2c_3 s_1^2s_2s_3\sin\delq\\
&J_{\rm PMNS}=-{\rm Im}\,[U_{31}U_{22}U_{13}]= C_1C_2C_3 S_1^2S_2S_3\sin\dell.\label{eq:JDet}
}

\begin{figure}[!t]
\begin{center}
\includegraphics[width=0.55\linewidth]{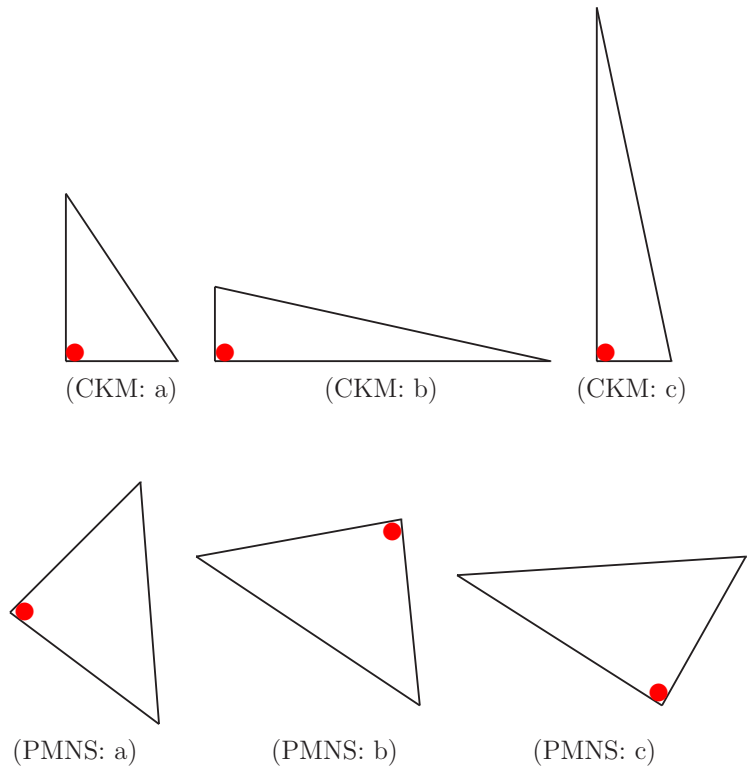}
\end{center}
\caption{Schematic shapes of $J$ for the CKM and PMNS triangles. All of them have one angle with $\delq=\pm\dell$ in our scenario.  } \label{fig:Js}
\end{figure}

For the neutrino masses, the heavy Majorana couplings are involved in addition. These latter couplings lead to large values of $|S_i|$, in contrast to small values of $|s_i|$. The Majorana couplings can introduce two more Majorana phases. These two Majorana phases $\delta_M$ cannot be measured independently from the Dirac phase in the laboratory experiments. A newly defined Dirac phase in the PMNS matrix is a combination from VEV $\langle X\rangle=V e^{i\delta}$ and $e^{-i\delta_M}$ of $M_N^{-1}$ of Fig. \ref{fig:MajNuMass}.  In this way, the phases $\delq$   and $\dell$  can be related. We assumed that the phase of $\langle X\rangle$ is the only source of CP violation, including the heavy neutrino sector. The mass matrix of $m_{ab}$ of Fig. \ref{fig:MajNuMass} is a combination of phases of $f^{\nu\,*}_{ab}$ and the phase $e^{-i \delta_M/2}$ which must define $\dell$.  If $\delta$ of  VEV $\langle X\rangle$ is zero, there is no CP violation in the quark sector, and also  in the lepton sector, \ie $\delq=0, \dell=0$, and $\delta_M=0$ in the full theory.  
We present a physical argument  to glimpse the situation without a detail study.  The Jarlsgog triangles in Fig. \ref{fig:Js} becomes lines and $J=0$ if $\delq$ and $\dell$ are integer multiples of $\pi$. Then, there is no physically measurable CP violation. Whatever happens in the calculation, this must be the case. There will be no weak CP violation if $\delta$ (the phase of $X$) is integer multiples of $\pi$ since the VEV of $X$ does not introduce an imaginary component in the whole theory.  Therefore, $\delq$ and $\dell$ musy be integer multiples of $\delta$ not to introduce CP violation in case $\delta=\pi$. This must be true even if we consider the heavy neutrinos since the heavy neutrinos belongs to a part in the theory.

\dis{
\delq=n_1\delta,~\dell= n_2 \delta . \label{eq:signorder}
}
 Because our argument on the vanishing of $J$ does not depend on the sign of $\delq$,  we can take both signs for $\delq$ and $\dell$, and for $n_1=n_2$ we have
\dis{
\dell=\pm\delq. \label{eq:Test}
} 
Note that we obtained this result by assuming that only the phase of $\langle X\rangle$  is the source of the weak CP violations, including the heavy neutrino sector. Namely, in the full theory the Majorana phases must also arise without coupling to $X$ or by the phase of $\langle X\rangle$. This physical argument does not depend on which parametrization we use for the PMNS matrix. Namely, we can use any parametrization for the PMNS matrix as far as $\dell$ is one of $\alpha,\beta,$ and $\gamma$ of PDG.\footnote{ 
We assume the unification of CP phases and the CKM phase can be one of  $\alpha,\beta,$ and $\gamma$ which are already determined from the O($\lambda^3)$ unitarity triangle.}
 In other words, an accurate determination of $\dell$, which will be the common angle appearing in all six leptonic Jarlskog triangles, will fix $\delq$ and  choose one class from the CKM matrices.\footnote{Note that $\delq$ is also the common angle  in all six quark Jarlskog triangles, but in each triangle except the one with angles $\alpha,\beta$ and $\gamma$ in the PDG book one side is always small, which makes it difficult to measure the angles at the end of the small side.}
 In Fig. \ref{fig:Js}, we show that one Jarlskog phase appears in the CKM and PMNS triangles if the assumptions on our CP violation are satisfied.   

%%%%%%%%%%%%%%%%%%
%%%%%%%%%%%%%%%%%%%%%

We followed the flipped-SU(5) language so far. We can present the same line of reasoning line by line for the Georgi-Glashow model also \cite{GG74}: all Yukawa couplings are real and the VEVs $\langle H_u\rangle $ and  $\langle H_d\rangle $ are real and   $\langle X\rangle$ obtaining a complex VEV couples only to charged leptons and $\Qem=-\frac13$ quarks. In this case, the GUT breaking is by a VEV of an adjoint BEH field $\langle{\bf 24}\rangle$. However, a dilemma here is the difficulty of obtaining the SU(5) adjoint BEH field from string compactification \cite{Tye96}. In addition, the doublet-triplet splitting with the adjoint BEH field needs a fine tuning between the VEV $\langle{\bf 24}\rangle$ and a free mass parameter for the adjoint scalar $m^2\,{\bf 24}^2$.

%%%%%%%%%%%%%%%%%%%%%%%%
\section{Relating the phase of $\langle X\rangle$ to $\delq$ and $\dell$}\label{sec:CCKMphase}

Even though the physical argument presented in the previous section is enough for relating $\dell$ and $\delq$, here we present a scheme in detail how they are connected.
The common phase in $\langle X\rangle=Ve^{i\delta}$ will appear in the $(ij)$ element of the quark or lepton mass matrix with a form, $ A_{ij}(e^{i\delta})^n$ if the singlet $X^n$ is located in the $(ij)$ element. We work in the bases where $M^{(d)}$ and $M^{(e)}$ are already diagonalized, and the left-hand unitary matrices diagonalizing $M^{(u)}$ and $M^{(\nu)}$ are the CKM and PMNS matrices, respectively. 

The parameters  of mass matrix $M^{(u)}$ combine to produce the phase in the CKM matrix.
  For a complex mass matrix $M^{(u)}$, there are 18 real parameters. It can be diagonalized by a bi-unitary transformation by $U^L$ and $U^R$. Thus, there are 36 parameters to be considered initially from $M^{(u)}, U^L$ and $U^R$. The left- and right-hand sides of the diagonalization relation,  $U^{L \dagger}M^{(u)} U^{R}=e^{i\alpha}$(real diagonal mass matrix), have the same overall phase, corresponding to the baryon number conservation. Disregarding the baryon number, by making Det.$M^{(u)} =$ real so that the diagonalized masses are real, we consider 35 real parameters in the diagonalizing conditions. The condition $U^{L \dagger}M^{(u)} U^{R}$=\,(real diagonal mass matrix) gives 18 relations. With these conditions imposed, then there remain 17 independent parameters from 35. Out of 17, five (since the overall phase cannot be used) can be removed by redefining L- and R-handed $u$-type quark phases. Thus, there remain 12 independent parameters. Out of 12, 11 parameters remain as physical ones, three real $u$-type quark masses, four angles of $U^L$, and  four angles of $U^R$. But, parameters in $U^R$ are hidden at low energy.\footnote{We choose the same number of parameters for $U^R$ as for $U^L$, since the same physics must result from quantum fields with (fields)$^c\leftrightarrow$(fields) which is equivalent to L$\leftrightarrow$R.}
    Now, there is one more (phase) parameter remaining. So, we must use one more relation to fix the theory completely. It is the relation $\delq=n_1 \delta$. 
  
  Let us parametrize  $U^{L,R}$ as given in (\ref{eq:KSCKM}) in which case there is no more freedom to rotate the quark fields,
  \dis{
 M_{\alpha\beta}= U^{L*}_{i\alpha}M^{(u)}_{ij}U^R_{j\beta},~
 \ie~
 M_{1\beta}= U^{L*}_{i1}M^{(u)}_{ij}U^{R}_{ j\beta}=U^{L*}_{i1}M^{(u)}_{i1}U^{R}_{13}+U^{L*}_{i1 }M^{(u)}_{i2}U^{R}_{23}+U^{L*}_{i1}M^{(u)}_{i3}U^{R}_{33}. \label{eq:Mij}
 }
 In (\ref{eq:KSCKM}), we try to fix the phase $\delq$ from the (31) element since there is no ambiguity in choosing the phase because it is an overall one, $\propto e^{i\delq}$. From (\ref{eq:Mij}), note that the (31) element appears only in $M_{1\beta}$: in the factor $U^{L*}_{i1}$. In the other elements, they are real or a phase does not appear as an overall one in Eq. (\ref{eq:KSCKM}). 

 \dis{
M_{13} =&U^{L*}_{11}M^{(u)}_{11}U^{R}_{13}+ U^{L*}_{21}M^{(u)}_{21} U^{R}_{13}+
\underline{U^{L*}_{3 1}M^{(u)}_{31}}U^{R}_{13} \\ 
 +&U^{L*}_{11}M^{(u)}_{12}U^{R}_{23}+U^{L*}_{21}M^{(u)}_{22}U^{R}_{23}+U^{L*}_{31 }M^{(u)}_{32}U^{R}_{23} \\
+ &U^{L*}_{11}M^{(u)}_{13}U^{R}_{33}+U^{L*}_{21}M^{(u)}_{23}U^{R}_{33}+U^{L*}_{31}
 M^{(u)}_{33}U^{R}_{33}=0,\label{eq:13}
 }
  \dis{
M_{12} =&U^{L*}_{11}M^{(u)}_{11}U^{R}_{12}+U^{L*}_{21}M^{(u)}_{21}U^{R}_{12}+\underline{U^{L*}_{3 1}M^{(u)}_{31}}U^{R}_{12} \\ 
 +&U^{L*}_{11}M^{(u)}_{12}U^{R}_{22}+U^{L*}_{21}M^{(u)}_{22}U^{R}_{22}+U^{L*}_{31 }M^{(u)}_{32}U^{R}_{22} \\
+ &U^{L*}_{11}M^{(u)}_{13}U^{R}_{32}+U^{L*}_{21}M^{(u)}_{23}U^{R}_{32}+U^{L*}_{31}
 M^{(u)}_{33}U^{R}_{32}=0,\label{eq:12}
 }
 \dis{
M_{11} =&U^{L*}_{11}M^{(u)}_{11}U^{R}_{11}+U^{L*}_{21}M^{(u)}_{21}U^{R}_{11}+
\underline{U^{L*}_{3 1}M^{(u)}_{31}}U^{R}_{11} \\ 
 +&U^{L*}_{11}M^{(u)}_{12}U^{R}_{21}+U^{L*}_{21}M^{(u)}_{22}U^{R}_{21}+U^{L*}_{31 }M^{(u)}_{32}U^{R}_{21} \\
+ &U^{L*}_{11}M^{(u)}_{13}U^{R}_{31}+U^{L*}_{21}M^{(u)}_{23}U^{R}_{31}+U^{L*}_{31}
 M^{(u)}_{33}U^{R}_{31}=m_u.\label{eq:11}
 }
Since $U^R_{1\alpha}$ is real in the KS form, 
we choose $\delq$ as the argument of $M^{(u)}_{31}$, 
making the underlined parts of Eqs. (\ref{eq:13},\ref{eq:12},\ref{eq:11}) real. The number of conditions in $M_{\alpha\beta}= U^{L*}_{i\alpha}M^{(u)}_{ij}U^R_{j\beta}$ is 18, which we counted before. We make this number to 19 by imposing an extra condition, $U^{L*}_{3 1}M^{(u)}_{31}=$\,real, where $M^{(u)}_{31}\propto \langle X\rangle\propto e^{i\alpha}$. This is a detail construction of $\delq$ from  $M^{(u)}$.

In the leptonic sector, consider  Fig. \ref{fig:MajNuMass}. We assumed that the charged lepton mass matrix $M^{(e)}$ is already diagonalized. The symmetric neutrino mass term is $\frac12\nu^T M^{(\nu)}\nu$, violating the lepton number by two units. Here, $M^{(\nu)}$  is complex and symmetric.  A complex symmetric matrix $A$ can be `diagonalized' using one unitary matrix $U$, where $U^T AU$  is a real diagonal matrix, which is called the Autonne-Takagi factorization \cite{ATfactorization}. It is not a unitary transformation,
\dis{
 M^{(\nu)\,\rm diag.}_{\alpha\beta} =& U_{i\alpha}M^{(\nu)}_{ij}U_{j\beta}.
 \label{eq:Mnuij}
  }
Even though the theory breaks the lepton number, the overall phase cannot be used in the diagonalizing condition (\ref{eq:Mnuij}), since both in the left- and right-hand sides break the lepton number by the same unit. Thus, the independent number of conditions in (\ref{eq:Mnuij}) is 17.
The (31) element appear  in both $M^{(\nu)\,\rm diag.}_{\alpha 1}$ and $M^{(\nu)\,\rm diag.}_{1\beta}$: in the factor $U_{i1}=U_{1i}$. In the other elements, they are real or a phase does not appear as an overall one in Eq. (\ref{eq:KSCKM}).  Note that $U_{31}$ appears in
 \dis{
 M_{1 1}=& U_{11}M^{(\nu)}_{11}U_{11}+U_{11 }M^{(\nu)}_{12}U_{21}+ \underline{U_{11}M^{(\nu)}_{13}U_{31}} \label{eq:P11}
\\
+& U_{21}M^{(\nu)}_{21}U_{11}+U_{21 }M^{(\nu)}_{22}U_{21}+ U_{21}M^{(\nu)}_{23}U_{31},
\\
+& \underline{U_{31}M^{(\nu)}_{31}U_{11}}+ U_{31 }M^{(\nu)}_{32}U_{21}+ U_{31}M^{(\nu)}_{33}U_{31}=m_{\nu_e},
 }
\dis{
 M_{2 1}=& U_{12}M^{(\nu)}_{11}U_{11}+U_{12 }M^{(\nu)}_{12}U_{21}+ \underline{ U_{12}M^{(\nu)}_{13}U_{31}}\label{eq:P21}
\\
+& U_{22}M^{(\nu)}_{21}U_{11}+U_{22 }M^{(\nu)}_{22}U_{21}+ U_{22}M^{(\nu)}_{23}U_{31},
\\
+& U_{32}M^{(\nu)}_{31}U_{11}+U_{32 }M^{(\nu)}_{32}U_{21}+ U_{32}M^{(\nu)}_{33}U_{31} =0,
 }
\dis{
 M_{3 1}=& U_{13}M^{(\nu)}_{11}U_{11}+U_{13 }M^{(\nu)}_{12}U_{21}+\underline{U_{13}M^{(\nu)}_{13}U_{31}} \label{eq:P31}
\\
+& U_{23}M^{(\nu)}_{21}U_{11}+U_{23 }M^{(\nu)}_{22}U_{21}+ U_{23}M^{(\nu)}_{23}U_{31},
\\
+& U_{33}M^{(\nu)}_{31}U_{11}+U_{33 }M^{(\nu)}_{32}U_{21}+ U_{33}M^{(\nu)}_{33}U_{31} =0,
 }
 \dis{
 M_{12}=& U_{11}M^{(\nu)}_{11}U_{12}+U_{11 }M^{(\nu)}_{12}U_{22}+U_{11}M^{(\nu)}_{13}U_{32}\label{eq:P12}\\ 
  +&U_{21}M^{(\nu)}_{21}U_{12}+U_{21 }M^{(\nu)}_{22}U_{22}+U_{21}M^{(\nu)}_{23}U_{32}\\
+& \underline{U_{31}M^{(\nu)}_{31}U_{12}}+ U_{31 }M^{(\nu)}_{32}U_{22} + U_{31}M^{(\nu)}_{33}U_{32} =0,
}
\dis{
 M_{13}=& U_{11}M^{(\nu)}_{11}U_{13}+U_{11 }M^{(\nu)}_{12}U_{23}+U_{11}M^{(\nu)}_{13}U_{33}\label{eq:P13}\\
 +& U_{21}M^{(\nu)}_{21}U_{13}+U_{21 }M^{(\nu)}_{22}U_{23}+U_{21}M^{(\nu)}_{23}U_{33}\\
+& \underline{U_{31}M^{(\nu)}_{31}U_{13}}+ U_{31 }M^{(\nu)}_{32}U_{23} + U_{31}M^{(\nu)}_{33}U_{33}  =0.
}
The number of conditions in (\ref{eq:P11},\ref{eq:P21},\ref{eq:P31},\ref{eq:P12},\ref{eq:P13})  is 10. But, we impose an additional condition $U_{31}M^{(\nu)}_{31}=U_{31}M^{(\nu)}_{13}=$\,real shown as underlined parts. 
We choose  $\dell$ as the argument of $M^{(\nu)*}_{31}=M^{(\nu)*}_{13}$.
Thus, the number of conditions we impose in (\ref{eq:P11},\ref{eq:P21},\ref{eq:P31},\ref{eq:P12},\ref{eq:P13}) is 11. Then, the total number of conditions we impose in $M^{(\nu)\,\rm diag.}_{\alpha\beta} =  U_{i\alpha}M^{(\nu)}_{ij}U_{j\beta}$ is $18=17+1$. The total number of parameters we introduced in $M^{(\nu)}$ and $U$ was $27=18+9$. Imposing 18 conditions, thus, there remain 9 physical parameters out of 27. These are three neutrino masses, two Majorana phases, and $\Theta_1,\Theta_2,\Theta_3$, and $\dell$ in the PMNS matrix. Thus, from our parametrization (\ref{eq:KSCKM}), we obtain
$\dell=-$(phase of $M^{(\nu)}_{31}$).  

In fact, in the anti-SU(7) model of \cite{KimJHEP15}, we can show this scheme. Since we have not obtained singlet representations yet, we cannot discuss two Majorana phases.\footnote{Singlet representations in the anti-SU(7) model will be presented in the future \cite{Kimfuture}.}
The nonsinglets in (\ref{eq:OneSU5family}) contain three neutral heavy leptons in three $\ten$'s. These are interpreted as $N$'s of Fig.  \ref{fig:MajNuMass}\,(c). We worked in the bases where $M^{(d)}$ and $M^{(e)}$ are diagonal. Note that the $\five_H$ couplings are a simplified version of $\five_H\cdot$(singlets) \cite{KimJHEP15,CKMUGUTF15}. Namely, the VEVs of the singlets multiplied with $\five_H$ are real. Therefore, masses of $N$ of Eq.  (\ref{eq:OneSU5family}), resulting from $\ten\cdot \ten\cdot \langle\five_H\rangle$, are real. Namely, the Majorana masses of $N$ in  Fig.  \ref{fig:MajNuMass}\,(c) are real,  \ie $(M^{(N)})^{-1}$ does not introduces a phase in  $M^{(\nu)}$.

In the flipped-SU(5) language, both $M^{(u)}$ and the Dirac mass in  $M^{(\nu)}$ appear from  $\ten \cdot\fiveb\cdot\fiveb_{H}$. Of course, $\fiveb_H$ couplings imply $\fiveb_H\cdot$(singlets) where some singlets contain $X^n$.  Namely, $M^{(u)}$ results  from $\ten({\rm containing}~u)\cdot\fiveb ({\rm containing}~u^c)\cdot\fiveb_{H}$ couplings and the Dirac couplings for $M^{(\nu)}$ appear from $\ten({\rm containing}~N)\cdot\fiveb({\rm containing}~\nu)\cdot\fiveb_{H}$. 
 But, our explicit calculation above needs only effective couplings. The leading term of the (31) element of the $u$-type quark mass matrix in the $\Z_{12-I}$ model \cite{KimJHEP15} takes the form
\dis{
 \bar{\bf 7}(T_3)_3\cdot{\bf 21}(U)_1\cdot \bar{\bf 7}_{\rm BEH}(T_6)\cdot  {\one}_{\rm BEH}(T_3)\label{eq:massu7}
 }
 where the subscripts are the family indices and the twisted sectors are $T_i$ and the untwisted sector is $U$.  The leading term of the (31) element of the neutrino mass matrix in the $\Z_{12-I}$ model \cite{KimJHEP15} takes the form
 \dis{
\bar{\bf 7}(T_3)_3\cdot\bar{\bf 7}(T_3)_1\cdot {\bf 7}_{\rm BEH}(T_3)\cdot {\bf 7}_{\rm BEH}(T_3)\cdot  {\one}_{\rm BEH}(T_3)\cdot  {\one}_{\rm BEH}(T_9). \label{eq:massnu7}
}
Remember that we chose $\delq$ as the argument of $M^{(u)}_{31}$ and  $\dell$ as the argument of $M^{(\nu)*}_{31}=M^{(\nu)*}_{13}$. In Eqs. (\ref{eq:massu7}) and  (\ref{eq:massnu7}), the only complex singlet is $ {\one}_{\rm BEH}(T_3)$. Thus, we obtain $\dell=-\delq$, realizing $n_2=-n_1$ of Eq. (\ref{eq:signorder}).

%%%%%%%%%%%%%%%%
\section{Conclusion}\label{sec:Conclusion}
  
A preliminary value for $\dell$ is large \cite{OyamaPlanck15}, posing a theoretical question whether $\delq=\pm\dell$ is satisfied or not. Thus, the relation between quark and lepton parameters, if true, must originate from a kind of GUT relation (for the quark and lepton parameters) in a families-unified model (to calculate the $\delq$ and $\dell$). We presented a possibility for this relation if the weak CP violation is of spontaneous origin {\`a la} Froggatt and Nielsen \cite{FN79} with only one complex VEV of a standard model singlet field $X$. Thus, proving this relation accurately hints a GUT, family unification, and spontaneous CP violation. In addition, an accurate determination of the Jarlskog phase in the PMNS matrix will pin down one class of the currently allowed CKM parametrizations by the relation  $\delq=\pm\dell$ from topological argument. We have shown that $\delq\simeq - \dell$ at the leading order if the GUT is the flipped-SU(5). 
      
 %%%%%%%%%%%%%%%%%%%%%%%%%%%%%%%%%%%%%%%%%%%%%%%%%%%%%%%%%%%%%%%%%%%%
\acknowledgments{We would like to thank Hyung Do Kim for useful communications. This work is supported in part by the National Research Foundation (NRF) grant funded by the Korean Government (MEST) (No. 2005-0093841) and by the IBS (IBS-R017-D1-2014-a00). }

%%%%%%%%%%%%%%%%%%
%%%%%%%%%%%%%%

  %%%%%%%%%%%%%%%%%%  

\end{document}